\documentclass[showpacs,amsmath,amssymb,preprintnumbers,twocolumn]{revtex4}

\usepackage{graphicx}
\usepackage{dcolumn}
\usepackage{bm}
\usepackage{amsmath}

\begin{document}
\preprint{APS/123-RED}

\title{Structural transition of force chains observed by mechanical spectroscopy}

\author{Wan-Jing Wang, Kai-Wei Yang, Xue-Bang Wu, Yu-Bing Wang, and Zhen-Gang Zhu}
\email{zgzhu@issp.ac.cn}
\affiliation{ Key Laboratory of Materials
Physics, Institute of Solid State Physics, Chinese Academy of
Sciences, P.O. Box 1129, Hefei, Anhui, People's Republic of China,
230031}

\date{\today}

\begin{abstract}
The dissipation properties of a fine sand system are investigated
by a low-frequency mechanical spectroscopy. The experiments show
many interesting profiles of the relative energy dissipation,
which imply that some structural transition of force chains in
dense granular media has occurred. The following data and
discussion indicate that the transition of force chains will lead
to the small deformation of arrangement in the granular system,
which is responsible for the historical effects. We hope this
research can improve our knowledge of the microstructure of the
granular materials.
\end{abstract}


\maketitle

\section{\label{sec:sec1}introduction}
Granular materials \cite{Gennes} are ubiquitous in everyday¡¯s
life, but a satisfying comprehension of their complex dynamics has
not yet been achieved. Force chains \cite{Liu1, Behringer} play a
key important role in understanding the static structure and
dynamics properties of the densely packed granular materials, such
as jamming \cite{Negel}, sound transmission \cite{Liu2}, force
propagation \cite{Goldenberg} and memory effects \cite{Geng}. Many
experiments and theoretical simulations have been carried out and
focused mainly on the characters of force chains
\cite{Coppersmith,Peters}. The most common method used to examine
force chains and theirs key characteristics is to visualize these
contact forces by stress-induced birefringence within assemblies
of photoelastic grains \cite{Liu1, Behringer}. Experiments and
simulations \cite{Negel, Radjai} all show that the transmission of
force chains appears as a complex force network that is highly
ramified and distributes inhomogeneous throughout the whole
system. In fact, the force chains are quite sensitive to small
perturbations in the packing geometry of the grains
\cite{Liu1,Liu2,Goldenberg,Brujic}. For example, if a granular
materials are driven by a small shear, the force chains will
change dramatically and evolve in a complex random way
\cite{Mueth, Anna, Goldenberg}. So recently an alternative method
has been developed to probe the force chains and its change
\cite{Anna1, Wang}. This method is based on the relationship
between the energy dissipation and the structural changes of force
chains in the granular system, which is related to the stress
relaxation process \cite{Brujic}.

In Ref. \cite{Anna1, Wang}, the dissipation properties of a
sheared granular medium have been studied and a simple rheological
model is presented, which suggests that small slides in the
inhomogeneous granular materials are responsible for the energy
dissipation. However, the microscopic picture about the slides is
unclear. In this work, our investigation will concentrate on the
details of the small slides in the microscopic picture, which we
call the structural transition of force chains. The mechanical
spectroscopy exhibits a fact that a small changes of the force
network structure in granular system can lead to a corresponding
energy dissipation. In addition, our experimental data show that
in the granular system the arrangement and the volume fraction of
the grains will change with the structural transition of force
chains. The following discussion indicates that it is the small
change of the arrangement that leads to the well-known historical
effects, which mean that the state and properties of granular
materials are quite dependent on their past history.

\begin{figure}[htbp]
\includegraphics[width =7.0cm]{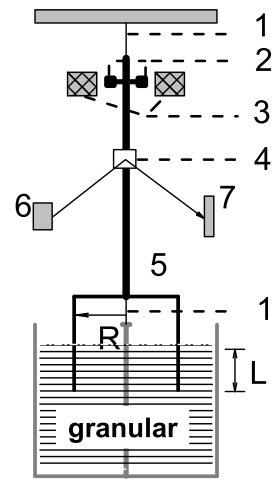}
\caption{\label{fig:Fig1} Sketch of the forced torsion pendulum
immersed into the granular medium. 1, suspension wires; 2,
permanent magnet; 3, external coils; 4, mirror; 5, cylinder (made
of aluminum alloy with inner radii R) un- or covered by a fixed
layer granules; 6, optic source; 7, optic detector}
\end{figure}
\section{\label{sec:sec2}experiment}
In the experiment, the mechanical spectroscopy measurements were
conducted on a developed low-frequency inverted torsion pendulum
using the forced-vibration method. Fig. \ref{fig:Fig1} shows the
sketch of the pendulum. The inverted torsion pendulum consists of
a cylinder being able to rotate around its axis, but prevented
from moving sideways by two suspension wires fixed to two ends of
the cylinder. The cylinder is forced into torsional vibration by a
time-dependent force $F(t)=F_0 \sin(2\pi \nu t)$, exerted by
applying a pair of permanent magnets fixed to the pendulum and
external coils (circulating an ac current), where $\nu$ is the
forced frequency. The angular displacement function of the
cylinder, $A(t)$, is measured optically. In the case here, the
response of the argument $A(t)=A_0 \sin(2\pi \nu t+
\delta\alpha)$, where $\delta\alpha$ is the phasic difference
between $A(t)$ and $F(t)$. According to this measurement
technique, the relative energy dissipation (RED) is calculated by
measuring the loss angle between applied stress and resulting
strain. Meanwhile, the relative modulus (RM) is calculated from
the ratio between the stress and strain.

Before each experiment, the granular system is flattened and
vibrated by external vibrations to ensure the accuracy of the
measurements. And the whole system is placed on an antivibrational
table to prevent undesired vibration-induced effects. In our
experiments the granular materials are composed of fine sand
sieved diameter $d=0.054\sim0.11mm$ and the maximum angular
displacement is below $0.4^{o}$ ($0.4\pi R/180<0.1d$, which
indicates that the real displacement of the cylinder is less than
$0.1d$, relative to the averaged particle size. So the system can
be considered as quasistatic.). The forced frequency $\nu=0.6 Hz$
or $\nu=1.0 Hz$ is chosen, which is well below the inherent
frequency of the pendulum (about $36 Hz$).

\section{\label{sec:sec3}results}
In our experiments the RED and the RM are given by
$\eta=\tan(\delta\alpha)$ and $G_0=F_0/A_0$, respectively. The
mechanical spectroscopy (RED and the RM) as a function of the
amplitude $A_0$, are shown as follow.

\begin{figure}[htbp]
\includegraphics[width=8.0cm]{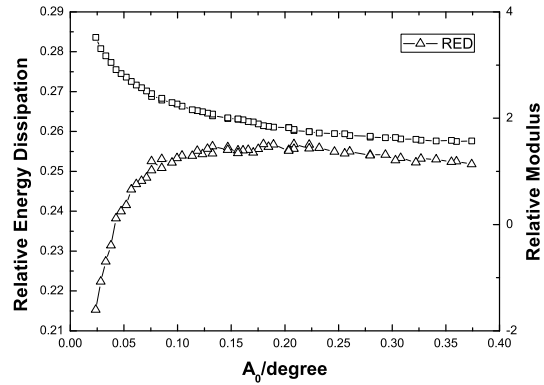}
\caption{\label{fig:Fig2} The RED ($\triangle$) and RM ($\square$)
virus the amplitude $A_{0}$(degree) in a fine sand system with the
oscillating cylinder covered by a layer of grains, for the
frequency $\nu=0.6 Hz$ and the immersed depth $h=60 d$.}
\end{figure}

\begin{figure}[htbp]
\includegraphics[width=8.0cm]{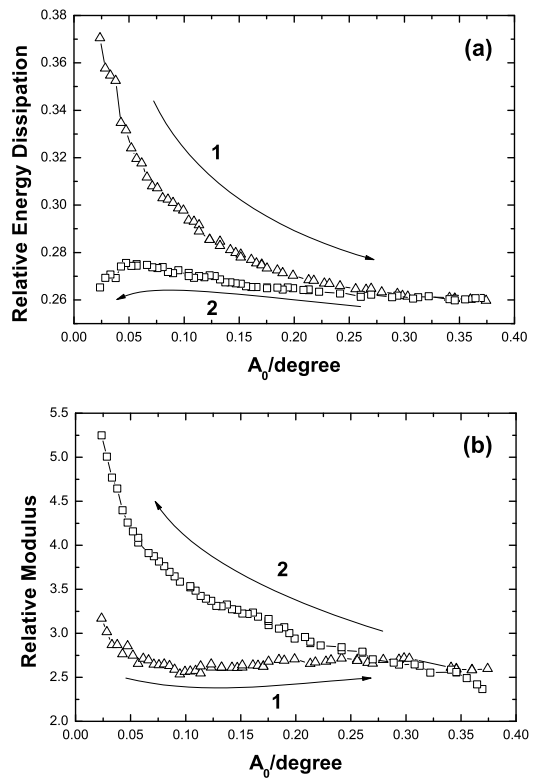}
\caption{\label{fig:Fig3} The historical effects of granular
materials. (a) The RED virus the amplitude with different path.
(b) RM virus the amplitude with different path. The immersed depth
$h=60 d$ and the forced frequency $\nu=0.6 Hz$.}
\end{figure}

Figure~\ref{fig:Fig2} shows clearly the relative energy
dissipation increases with the amplitude. Meanwhile, we observe
that the relative modulus decreases monotonously with increasing
amplitude. These data are obtained when we swept the amplitude
down repeatedly. Whether there is hysteresis in the system, i.e.
are the peaks still there if the forced amplitude is swept up and
down? This is an important question and relates to the historical
effects have been found in weakly vibrated granular systems
\cite{Umbanhowar}. As shown in Fig. \ref{fig:Fig3}(a), when we
increase the amplitude step by step, we obtain a slowly decreasing
RED. In succession, when the amplitude is swept down slowly, two
different profiles (RED and RM) are obtained, as the profile (2)
in Fig. \ref{fig:Fig3} shows. We find that the differences between
them are obvious: a rather small RED is obtained and there is a
loss peak, while RM increases monotonously with decreasing
amplitude. This is the historical effects, which indicate that the
state and properties of granular materials are quite dependent on
their past history.

\section{\label{sec:sec4}structural transition of force chains}
In Ref.~\cite{Anna1,Wang}, a simple rheological model is presented
to reproduce the RED measured in granular materials. In the model,
the granular medium is characterized by slide unit and spring
unit. The model suggests that small slides in the inhomogeneous
granular materials are responsible for the energy dissipation.
However, the microscopic picture about the slides is unclear.
Below we will try to give some details of the small slides in the
microscopic picture and then give a explanation of the historical
effects.

\begin{figure}[htbp]
\includegraphics[width=9.0cm]{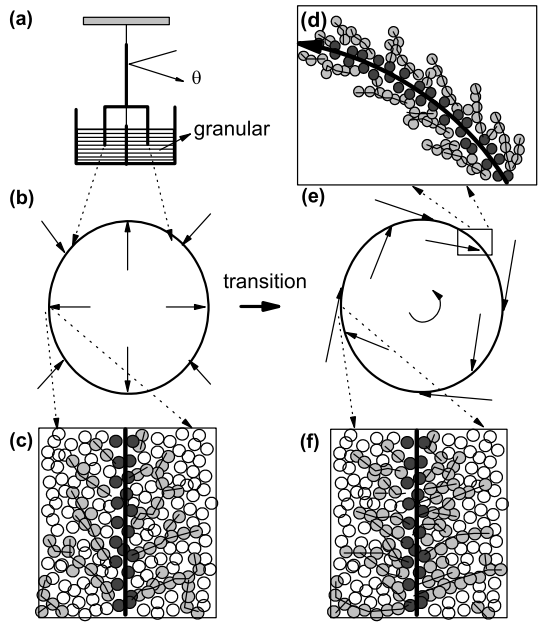}
\caption{\label{fig:Fig4} The structural transition of force
chains. (a) Sketch of the forced torsion pendulum immersed into
the granular medium. The cylinder is covered by a layer sand glued
on with epoxy. (b) Schematic (horizontal direction) and (c) the
microscopic picture (vertical direction) of the distribution of
force chains in a steady state. (e) Schematic (horizontal
direction) and (f) the microscopic picture (vertical direction) of
the distribution of force chains opposing the rotation of the
pendulum. The dark grains are glued on the cylinder by an epoxy.
The force chains are represented as bonds connecting the light
grey grains. Picture (d) shows the force chains in a horizontal
layer near the cylinder. In this picture, the grains besides those
on the force chains have not been shown.}
\end{figure}

Granular materials are relatively discrete medium and force
transmission through a granular system can only occur via the
interparticle contacts. Under the gravity force, the force
distribution in a static packing of grains in a cylinder should be
inhomogeneous and could form many force chains \cite{Geng,Radjai}.
Near the cylinder, the spatial force distribution will be
organized along directions almost normal to the cylindrical probe
\cite{Anna1}, where maximum strain be built up between chains of
grains, due to the gravity. These grain chains of forces have two
characters: first, the contact force between them is rather
stronger and carries almost all the weight load of the above
grains, and second, a chain is a quasilinear arrangement of three
or more grains \cite{Peters}. The chain lengths are defined as the
grain number in the chain, e.g., the shortest chain is two-grain
chain. Meanwhile the magnitude of the contact force denotes the
correlation strength of the chain. Here we present a sketch of the
grain chains in Fig. \ref{fig:Fig4}(b) and (c).
Fig.\ref{fig:Fig4}(c) shows a part of the microscopic picture
(vertical direction) of the distribution of force chains in static
state. As this picture shows, these force chains are oriented
almost in the vertical direction because of the weight of the
grains. In other words, the forces originate from the weight of
the grains and the weight of an above particle is transmitted to a
neighbor underlying particle, as shown in q-model \cite{Liu1,
Coppersmith}. So we often find the force chains arrange as roots
of trees. In addition, we know that the qualities and strengths of
chains will increase with the depth (``Janssen effects").

In the granular systems, the grains can not move independently.
More often some degree of freedom of a grain are partially frozen,
so that the motion of the grains is a correlated motion. When a
shear stress is applied to the granular material, rather than
deforming uniformly, the system such as dry sand develop shear
bands \cite{Drake} --- narrow zones of strongly correlated
particles, with essentially rigid adjacent regions. Similarly,
when a cylindrical probe is rotating in the granular medium, the
spatial force distribution around the probe will be organized
along directions almost tangent to the cylinder, where the maximum
stress also build up many chains of grains. These correlations of
the grain chains have been shown by the radial profiles of
azimuthal velocity in Ref. \cite{Mueth, Cates}. Here we present a
sketch of the grain chains under shear stress in Fig.
\ref{fig:Fig4} (e) and (f). Compared the conditions before the
shear, many differences will appear as follow. First, under shear
stress more neighbor grains of the cylindrical probe will join in
the force chains, although, the configuration of grains could not
change, as Fig. \ref{fig:Fig4} (c) and (f) show. Second, the
directions of force chains will change a lot, i.e., the force
chains oriented almost in the horizontal direction in contrast to
almost in the vertical direction. The reason is that the shear
stress changes the origin of the force chains. Then the shear
stress leads to the structural changes of the force chains. Except
for changing the arrangement of the force chains, the applied
shear will also influence the characters of the force chains,
e.g., the chain lengths will get longer and the correlation
strength of the chains will become stronger. In order to clearly
show the schematic of force chains, we give a microscopic picture
of the force chains in a horizontal layer near the cylinder as
shown in Fig. \ref{fig:Fig4} (d), where the other grains besides
the grains on the force chains have not been shown.

Here we call this change of the force chains without
configurational change as the structural transition of force
chains. Now let us discuss the relationship between the structural
transition of force chains and the energy dissipation. The
definition of force chains indicates that the drag force resisting
a solid object moving slowly through a granular medium originates
not only in the grains immediately in front of the object but also
in the successive layers of grains supporting them. When a
cylindrical probe begin to rotate in the granular medium, the
spatial force distribution around the probe will change
dramatically, as shown above. During this structural transition of
force chains, some change of configuration in microscopic length
scale, such as the formation and break of adhesive junctions
between the surface asperities, and other forms of localized
dissipative processes, must have occurred \cite{Anna1}.
Figure~\ref{fig:Fig2} shows clearly the relative energy
dissipation in these processes. When we increase the vibration
amplitude, the structural transition, such as the break of
adhesive junctions between the surface asperities, will increase,
i.e., the transition quantity will increase. When the transition
quantity increases the energy dissipation must will increase. As
expected, the experimental results also show that the energy
dissipation increases with the amplitude. Meanwhile, we observe
that the relative modulus decreases monotonously with increasing
amplitude, which indicates the soften of the dense granular system
under shear strain \cite{Wang}. These analysis indicate that the
structural transition of force chains leads to the energy
dissipation and the dissipation will increase with the transition
quantity.

In our experiments, the granular materials are composed of fine
sand sieved diameter $d=0.054\sim0.11mm$. In this scale, granular
matter is a well-known example of athermal system, that is a
system where classical thermodynamics does not apply since thermal
energy ($k_{B}T$) is insignificant compared to the gravitational
energy of a macroscopic grain. A static packing of grains is
therefore in a metastable state, indefinitely trapped in a local
minimum of the total potential energy. However, the forced
vibration will break the jamming of granular packing \cite{Anna}.
From Ref. \cite{Philippe} we know that after a succession of high
amplitude vibration a rather looser packing of grains (the volume
fraction is small) is obtained. However, we do not know what about
the mechanical spectroscopy in such a looser packing of grains
under low-amplitude shear. And we do not know how the volume
fraction of grains influences the energy dissipation of granular
system. In the following we will discuss the energy dissipation in
the condition of historical effects. As we know, the data in Fig.
\ref{fig:Fig2} are obtained in the experiments performed in the
direction that the forced amplitude is swept down repeatedly. In
the process we decrease the amplitude step by step and we obtain a
slowly decreasing RED. The beginning of the profile (1) in Fig.
\ref{fig:Fig3} shows the dissipation at low amplitude obtained
immediately after a succession of high amplitude vibration. In
other words, at this moment the granular packing is rather looser.
But at the beginning a rather larger RED and a rather lower RM are
obtained. Then Fig. \ref{fig:Fig3}(a) shows that the RED decreases
monotonously with increasing amplitude. In succession, when the
amplitude is swept down slowly, a denser packing is obtained
\cite{Philippe} and two different profiles (RED and RM) are
obtained, as the profile (2) in Fig. \ref{fig:Fig3} shows that a
rather small RED is obtained, while RM increases monotonously with
decreasing amplitude (see Fig. \ref{fig:Fig3}(b)). The above
discussion implies that the energy dissipation is related to the
volume fraction of granular system, i.e., the looser packing is
more dissipative. Considering that the energy dissipation of
granular system increases with the transition quantity of force
chains, we can say that the structural transition will occur more
easily in the looser packing.

\begin{figure}[htbp]
\includegraphics[width=8.0cm]{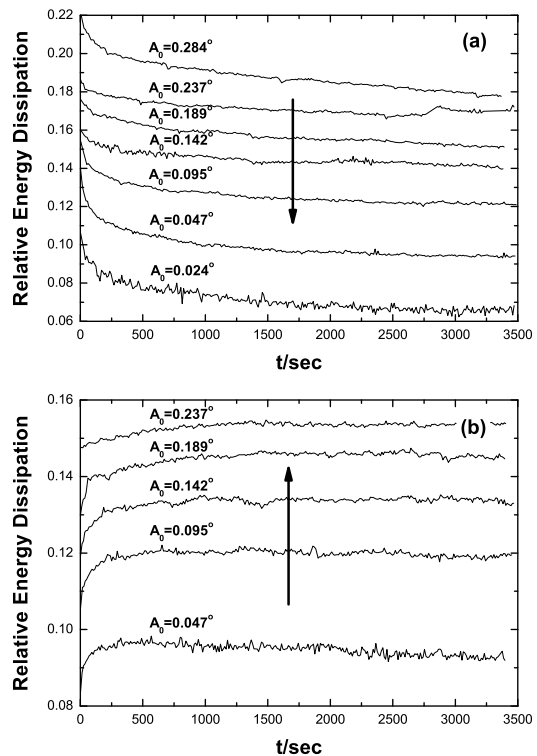}
\caption{\label{fig:Fig5} The RED virus the time with different
amplitude $A_0$, as noted. Firstly the amplitude is swept down
(a), and then the amplitude is swept up (b). The immersed depth
$h=70 d$ and the forced frequency $\nu=1.0 Hz$.}
\end{figure}

In order to understand the historical effects better, below we
will discuss the aging effects in granular materials. Fig.
\ref{fig:Fig5} shows the RED as a function of the time with
different amplitude $A_0$, as noted. At the beginning of all these
experiments, the packing is rather looser. Then under a series of
vibrations with the amplitude $A_0=0.284^{o}$, the granular system
will be compacted \cite{Philippe}. According to the relationship
between the volume fraction and the energy dissipation of granular
system, we know that when the granular system is compacted the RED
will decrease with the vibration time. As expected, Fig.
\ref{fig:Fig5}(a) shows the results: firstly the RED decreases
quickly and then slowly approaches a saturation value. In other
words, at the beginning the structural transition of force chains
will occur continuously with the energy dissipation of system.
Then the transition quantity will decrease with the compaction of
the granular packing, which is the reason of the decrease of
energy dissipation. These analysis indicate that the structural
transition of force chains has slowly changed the arrangement of
the granular materials, i.e., the compaction of granular
materials. And it is this arrangement change that leads to the
change of energy dissipation with vibration time. The profiles in
Fig. \ref{fig:Fig5}(a) can be well fitted with an exponential
decay law, which is a fundamental character of the relaxation of
granular systems \cite{Philippe}. This is the aging effects. Fig.
\ref{fig:Fig5}(a) presents many similar profiles with different
amplitude. The difference between them is that the quantity of RED
decreases with decreasing forced amplitude. However, when we
increase the amplitude step by step, we obtain some different
profiles. As Fig. \ref{fig:Fig5}(b) shows, at the beginning the
RED increases quickly and then slowly approaches a saturation
quantity, what indicates that the relaxation of granular system
has occurred and the arrangement of granular system must have
changed. As mentioned above, it also is the structural transition
of the force chains that leads to this change. However, the
profiles in Fig. \ref{fig:Fig5}(b) follow the exponential grow
law, which shows that the trend of the profiles is opposite with
the profiles in Fig. \ref{fig:Fig5}(a). These differences between
the profiles indicate that the arrangement of the grain packing
changes differently when we increase the amplitude step by step.
This is a process that the granular packing changes looser, in
contrast to the compaction. In addition, the results confirms an
earlier experimental observation: as the intensity of vibration
decreases both the volume fraction of stationary and the
compaction time increases \cite{Philippe}. The reason is that the
transition quantity will be very small when the amplitude of
vibration is lower. In according to the above analysis, we know
that the forced vibration actually changes the microstructure (the
volume fraction and the arrangement) of the granular system
because of the slowly structural transition of force chains. This
change of the microstructure is associated with the distribution
of the various-size empty apace between grains, i.e., the defects
motion and annihilation, which is the fundament of the historical
effects.

The above discussion indicates that the transition of force chains
is a relaxation process companied with the energy dissipation, as
shown in the explanation of experimental data. The mechanical
spectroscopy also show that a minute change of the arrangement of
the granular system is sufficient to significantly change the
amplitude response. It is these changes of the arrangement that
are responsible for the well-known historical effects in the
granular system preparation.

\section{\label{sec:sec5}conclusion}

No body will be surprised that when he moves a solid intruder in a
granular medium some deformations and some shear bands near
interface occur \cite{Gennes,Drake}. And we all know that the
shear bands are the key to explain the ``flowability" of granular
materials. However, the shear banding behavior is very complex and
is difficult to study. Many experiments and theoretical
simulations have carried out and shown that the shear band
thickness and shape are dependent on the shear strain and the
boundary roughness conditions \cite{Goldenberg,Siavoshi}. Here we
investigated the structural transition of force chains in the
shear bands from a view of energy dissipation. We found that the
shape of the shear bands will change with the shear stress, as the
microstructure of force chains shows. Of course, there are also
many questions in understanding what happens on the grain scale
\cite{Knight}.

Different from the most common method to study force chains using
birefringent materials \cite{Liu1, Behringer}, our investigation
focus on the relationship between the energy dissipation and the
force chains in granular systems, which gives a microscopic
picture of the force chains in dense granular materials. In the
experiments, we observed many interesting physical effects, such
as the historical effects and aging effects. The results indicate
that some structural transition of force chains, even changes of
arrangement, have occurred. While the energy dissipation in this
regime is a dynamic quality, we find that it is determined by the
static structure of the medium. So we can say this experiment
offers a new bridge between recent developments in understanding
static configuration in granular media and the dynamic properties.

This work is financially supported by the National Natural
Foundation of China under Grant No. 10674135 and the Knowledge
Innovation Program of Chinese Academy of Sciences under Grant No.
KJCX2-SW-W17.


\end{document}